\documentclass[11pt, a4paper]{article}

\usepackage[margin=1in]{geometry}
\usepackage{amsmath, amssymb}
\usepackage{graphicx}
\usepackage{hyperref}
\usepackage{booktabs}

\title{The Sheaf Laplacian: A Topological Framework for Data Fusion and Consensus in Distributed Sensing Networks}
\author{Manuel Hern\'andez, Eduardo S\'anchez-Soto}
\date{\today}

\begin{document}

\maketitle

\begin{abstract}
We argue here that traditional network models, which are overwhelmingly based on the mathematical construct of a simple graph, are fundamentally insufficient for capturing the complexity of modern distributed systems. Such systems are characterized by heterogeneous agents with diverse capabilities, high-dimensional and multi-modal data streams, and intricate, context-dependent relationships that cannot be adequately described by a simple connection or a scalar weight. The limitations of these classical models necessitate a new mathematical language, one with far greater expressive power. We have found that sheaf theory provides us with such a language. Moreover, we show that the sheaf Laplacian is a suitable mechanism for data fusion and establishing consensus within distributed sensing networks.
\end{abstract}

\section{Introduction}
The proliferation of distributed systems represents a defining technological shift of the 21st century. Networks of interconnected devices, ranging from vast wireless sensor networks (WSNs) for environmental monitoring and military surveillance to the burgeoning Internet of Things (IoT), autonomous drone swarms, and heterogeneous robotic teams, are generating data at an unprecedented scale \cite{bodnar2022neural, ghrist2022distributed, cassara2024cellular, yin2024lightweight}. In this new era, the fundamental challenge has evolved from mere data collection to the intelligent, decentralized integration and coordination of information. The core task is to transform a cacophony of local, noisy, and often conflicting measurements into a coherent global understanding upon which critical decisions can be made \cite{anandkumar2010distributed, ghrist2022distributed}.

This report advances the thesis that traditional network models, which are overwhelmingly based on the mathematical construct of a simple graph, are fundamentally insufficient for capturing the complexity of modern distributed systems. Such systems are characterized by heterogeneous agents with diverse capabilities, high-dimensional and multi-modal data streams, and intricate, context-dependent relationships that cannot be adequately described by a simple connection or a scalar weight \cite{hansen2021laplacians, ayzenberg2024sheaf}. The limitations of these classical models necessitate a new mathematical language, one with far greater expressive power. 

This report details such a new paradigm, drawn from the fields of applied topology and topological data analysis: the sheaf Laplacian \cite{hansen2021laplacians, hansen2019learning, purvine2018sheaf}. The sheaf Laplacian provides a canonical and deeply expressive language for modeling the local-to-global relationships inherent in distributed data \cite{robinson2016sheaves, hansen2021laplacians}. By leveraging the mathematical theory of sheaves, this operator offers a unified and principled approach to solving long-standing problems in data fusion, consensus, and signal processing. We will systematically build the case for the sheaf Laplacian, starting from the foundational challenges of distributed sensing, introducing the necessary mathematical machinery, and culminating in a discussion of its advanced applications and future horizons.

\section{The Landscape of Distributed Data Integration}
To appreciate the utility of the sheaf Laplacian, one must first understand the demanding environment in which modern sensor networks operate and the core computational problems they aim to solve. These networks are not idealized constructs but complex systems subject to a host of real-world constraints and failures.

\subsection{Characterizing the Distributed Sensing Environment}
Distributed sensing networks are defined by their inherent limitations and vulnerabilities. From a physical and logistical standpoint, individual sensor nodes are often constrained by limited communication bandwidth, finite energy reserves (battery power), and significant latencies, especially in large-scale or mobile deployments \cite{ghrist2022distributed, kar2007distributed}. Communication links themselves are frequently unreliable, subject to packet loss, intermittent failures, and the dynamic topology changes that occur when nodes move or go offline \cite{kar2008distributed, kar2007distributed}.

Beyond these operational constraints, the data itself is imperfect. Sensors are vulnerable to a wide range of environmental factors, leading to noisy or incorrect readings. The hardware is prone to defects, and in many applications, the network may be subject to malicious or adversarial attacks designed to inject misleading information \cite{cassara2024cellular, anandkumar2010distributed, kar2008distributed}. These issues---noise, faults, and attacks---are not rare exceptions but are defining characteristics of the operational environment that any robust data integration framework must address.

\subsection{The Core Problems: Data Fusion and Consensus}
Given this challenging environment, two central problems emerge: data fusion and consensus.

\textbf{Data Fusion} is the process of combining data and information from multiple, often heterogeneous, sources to produce an estimate or understanding that is more consistent, accurate, and useful than that provided by any single source \cite{barbarossa2020running, purvine2018sheaf}. In a sophisticated sensor network, this involves more than simply averaging numbers. It requires the fusion of different \textit{types} of data---for example, combining quantitative temperature readings with qualitative image classifications---from sensors that may have vastly different characteristics and error models \cite{purvine2018sheaf}. The ultimate goal is to synthesize these disparate pieces of local information into a single, coherent global picture.

\textbf{Consensus} is the process by which a group of distributed agents or nodes reach an agreement on a certain quantity of interest based solely on local interactions \cite{ghrist2022distributed, yin2024lightweight}. The canonical example is the ``average consensus'' problem, where each node begins with a private scalar value and, through iterative communication with its neighbors, converges to the global average of all initial values \cite{kar2008distributed, xiao2005local}. This serves as a foundational model for distributed coordination and information aggregation.

\subsection{A Review of Traditional Paradigms}
A variety of approaches have been developed to tackle these problems, each with its own strengths and weaknesses.
\begin{itemize}
    \item \textbf{Gossip Algorithms:} These are fully decentralized algorithms where nodes iteratively exchange information with their immediate neighbors to converge to a global value, such as the average. Protocols like Randomized Gossip and Push-Sum are popular due to their simplicity and robustness to dynamic topology changes, but they typically focus on aggregating simple scalar data \cite{cassara2024cellular, barbarossa2020running}.
    \item \textbf{Bayesian and Statistical Methods:} This class of methods has a long and successful history in signal processing. It includes techniques like likelihood ratio tests for optimal decision fusion and the Kalman filter for state estimation and tracking \cite{purvine2018sheaf}. These methods are powerful for handling noise and modeling system dynamics, but they often assume a centralized fusion center where all data is collected or rely on strong, and sometimes unrealistic, assumptions about the underlying statistical models of the system and noise.
    \item \textbf{Consensus Protocols:} Originating from the distributed computing community, protocols like Practical Byzantine Fault Tolerance (PBFT) and RAFT are designed to achieve agreement on a discrete state (e.g., the order of transactions in a blockchain) even in the presence of malicious (Byzantine) or crash failures \cite{yin2024lightweight, wang2024survey}. While extremely robust, they often incur high communication overhead and are not naturally suited for the fusion of continuous, high-dimensional data that is characteristic of sensor networks.
\end{itemize}

A careful review of these traditional methods reveals a fundamental dichotomy. On one hand, approaches from signal processing, such as the Kalman filter, are adept at handling noisy, continuous, vector-valued data but often presume a centralized architecture or a well-defined, linear system model \cite{purvine2018sheaf}. On the other hand, approaches from distributed computing, like gossip algorithms, are inherently decentralized and robust to network changes but typically oversimplify the data to scalar values or discrete states \cite{cassara2024cellular, yin2024lightweight}. This creates a significant gap: a need for a framework that is both fully distributed \textit{and} natively capable of handling the rich, high-dimensional, and heterogeneous data found in modern sensor networks.

Furthermore, the concept of ``running consensus'' highlights the dynamic nature of the problem. In this more realistic paradigm, nodes must simultaneously reach a consensus with their neighbors while continuously incorporating new local measurements from their environment \cite{barbarossa2020running}. This is not a static problem of averaging a fixed set of values but a dynamic process of continuous integration. The need to ``coherently modify'' the consensus equation to accommodate these new ``innovations'' calls for a more expressive mathematical operator than those used in standard models---one that can gracefully and structurally integrate new local information into an evolving global state. This sets the stage for the introduction of the sheaf Laplacian, which, through the dynamics it generates, provides precisely this mechanism.

\section{The Graph Laplacian: A Baseline for Network Dynamics}
Before delving into the complexities of sheaf theory, it is essential to understand the standard tool for analyzing dynamics and consensus on networks: the graph Laplacian. This matrix serves as a crucial baseline, and its limitations directly motivate the need for a more advanced framework.

\subsection{Formal Definition and Construction}
A network's communication topology is most commonly modeled as a graph, $G = (V, E)$, where $V$ is the set of $n$ nodes (sensors) and $E$ is the set of edges representing communication links \cite{numberanalytics2025laplacian, von2008tutorial}. The structure of this graph can be encoded in two fundamental matrices:
\begin{itemize}
    \item The \textbf{Adjacency Matrix} $A$, an $n \times n$ matrix where $A_{ij} = 1$ if an edge exists between nodes $i$ and $j$, and $0$ otherwise. For weighted graphs, $A_{ij}$ can represent the strength of the connection \cite{numberanalytics2025laplacian, mahdi2024graph}.
    \item The \textbf{Degree Matrix} $D$, a diagonal matrix where the entry $D_{ii}$ is the degree of node $i$ (the number of edges connected to it) \cite{numberanalytics2025laplacian, mahdi2024graph}.
\end{itemize}

The \textbf{Graph Laplacian} matrix, denoted $L$, is formally defined as the difference between the degree matrix and the adjacency matrix \cite{numberanalytics2025laplacian, wikipedia2024laplacian, mahdi2024graph}:
$$L = D - A$$
For example, for a simple graph with 3 nodes where node 1 is connected to nodes 2 and 3, the matrices would be:
$$
A = \begin{bmatrix} 0 & 1 & 1 \\ 1 & 0 & 0 \\ 1 & 0 & 0 \end{bmatrix}, \quad
D = \begin{bmatrix} 2 & 0 & 0 \\ 0 & 1 & 0 \\ 0 & 0 & 1 \end{bmatrix}, \quad
L = D - A = \begin{bmatrix} 2 & -1 & -1 \\ -1 & 1 & 0 \\ -1 & 0 & 1 \end{bmatrix}
$$

\subsection{The Laplacian Spectrum and Consensus}
The graph Laplacian has several key mathematical properties that make it central to network analysis. It is a symmetric and positive semi-definite matrix, which means all of its eigenvalues are real and non-negative: $0 = \lambda_1 \le \lambda_2 \le \dots \le \lambda_n$ \cite{numberanalytics2025laplacian, wikipedia2024laplacian, von2008tutorial}. The spectrum of this matrix---its eigenvalues and corresponding eigenvectors---reveals profound information about the graph's structure. Most importantly, the multiplicity of the eigenvalue $\lambda=0$ is equal to the number of connected components in the graph \cite{numberanalytics2025laplacian, wikipedia2024laplacian, von2008tutorial}. For a fully connected graph, there is only one zero eigenvalue. The corresponding eigenvector, which spans the kernel (or null space) of $L$, is the constant vector $\mathbf{1} = (1, 1, \dots, 1)^T$ \cite{numberanalytics2025laplacian, wikipedia2024laplacian}.

This property links the Laplacian directly to the problem of average consensus. The graph diffusion equation, often called the heat equation on a graph, is given by:
$$\frac{d\mathbf{x}}{dt} = -L\mathbf{x}$$ 
where $\mathbf{x}(t)$ is a vector of scalar values at each node at time $t$. This equation describes a process where each node's value changes based on the differences with its neighbors. The stable equilibria of this system are the states where $d\mathbf{x}/dt = 0$, which occurs if and only if $\mathbf{x}$ is in the kernel of $L$. For a connected graph, this means the system is stable only when $\mathbf{x}$ is a constant vector---that is, when all nodes have the same value. The system will converge to a state where each node's value is the average of the initial values across the network \cite{xiao2005local, numberanalytics2025laplacian}. Thus, the graph Laplacian is the canonical operator for modeling and achieving simple average consensus.

\subsection{Fundamental Limitations of the Graph Laplacian}
The very elegance and perfect suitability of the graph Laplacian for the average consensus problem became, in a sense, a conceptual constraint. Its success cemented the popular notion of ``consensus'' as this simple scalar agreement model, overshadowing the need to explore more complex forms of agreement for many years. The limitations of the graph Laplacian are not merely technical shortcomings but define the boundaries of this classical paradigm \cite{hansen2021laplacians}.
\begin{enumerate}
    \item \textbf{Scalar Data Assumption:} The operator $L$ is fundamentally designed for signals where each node $v$ holds a single scalar value $x_v$. It has no native mechanism for handling vector-valued data, functions, or other structured data types that are common in modern sensor systems \cite{hansen2021laplacians}.
    \item \textbf{Oversimplified Relationships:} The Laplacian only encodes the existence and, in weighted versions, the intensity of a connection. It cannot capture the \textit{nature} or \textit{type} of the relationship. For example, it cannot distinguish between a relationship of agreement and one of opposition, nor can it model more complex data transformations between nodes \cite{hansen2021laplacians, ayzenberg2024sheaf}.
    \item \textbf{A Narrow Definition of Consensus:} The only ``consensus'' state that the graph Laplacian can represent is one of perfect uniformity, where all nodes hold the same value. It is incapable of modeling more sophisticated consistency conditions, such as states where nodes must maintain specific ratios, alignments, or other structured, non-uniform relationships \cite{hansen2021laplacians}.
    \item \textbf{Inability to Model Heterogeneity:} The framework implicitly assumes a homogeneous network where all nodes and relationships are of the same kind. It struggles to represent networks composed of heterogeneous agents with different roles, capabilities, or internal state-space dimensions \cite{cassara2024cellular}.
    \item \textbf{Computational Expense:} While a practical rather than conceptual issue, computing the full spectrum of the Laplacian for very large networks can be computationally expensive, a challenge that is inherited and often amplified by more complex successors \cite{numberanalytics2025limitations}.
\end{enumerate}

The quadratic form of the Laplacian, $\mathbf{x}^T L \mathbf{x} = \sum_{(i,j) \in E} w_{ij}(x_i - x_j)^2$, provides a powerful physical intuition. It measures the total ``energy'' or ``tension'' of a signal on the graph as the sum of squared disagreements across all edges \cite{numberanalytics2025laplacian}. A state of consensus is a state of minimum energy. This core idea---quantifying global inconsistency as a sum of local tensions---is a foundational concept that sheaf theory will generalize profoundly, replacing the simple scalar difference $(x_i - x_j)$ with a much richer and more expressive measure of disagreement \cite{ayzenberg2024sheaf, mahdi2024moving}.

\section{A Primer on Cellular Sheaves for Network Modeling}
To overcome the limitations of the graph Laplacian, a more powerful mathematical structure is required. This structure is the \textbf{sheaf}, a concept from algebraic topology and geometry that provides the canonical framework for attaching data to spaces and tracking it in a systematic, locally-consistent manner \cite{robinson2016sheaves, bredon2012sheaf, tennison1975sheaf}. For applications in networks and data science, a discrete and computationally tractable version known as a \textbf{cellular sheaf} is used \cite{hansen2019distributed, ghrist2011computational}. The fundamental idea is to elevate a graph from a simple diagram of connections to a scaffold for structured data, where the nodes and edges themselves become containers for complex information \cite{ayzenberg2024sheaf, hansen2021laplacians}.

\subsection{The Anatomy of a (Cellular) Sheaf $\mathcal{F}$}
A cellular sheaf $\mathcal{F}$ over a graph $G=(V,E)$ is defined by two primary components: stalks and restriction maps \cite{hansen2019gentle}.
\begin{itemize}
    \item \textbf{Stalks:} These are the data spaces assigned to the individual elements (vertices and edges) of the graph.
    \begin{itemize}
        \item \textbf{Vertex Stalks $\mathcal{F}(v)$:} For each vertex $v \in V$, we assign a vector space $\mathcal{F}(v)$. This space holds the local data associated with that node. For a sensor, this could be the space of all possible measurements it can make; for an agent, it could be its internal state space \cite{ayzenberg2024sheaf, hansen2019distributed, hansen2021opinion}.
        \item \textbf{Edge Stalks $\mathcal{F}(e)$:} For each edge $e \in E$, we assign another vector space $\mathcal{F}(e)$. This space serves as a common frame of reference or a ``discourse space'' for the two nodes connected by the edge. It is the context in which their respective data can be compared \cite{ayzenberg2024sheaf, hansen2019distributed, hansen2021opinion}.
    \end{itemize}
    \item \textbf{Restriction Maps:} These are linear transformations that link the data across the network. For each edge $e=(u,v)$, there are two restriction maps, $\mathcal{F}_{u \to e}: \mathcal{F}(u) \to \mathcal{F}(e)$ and $\mathcal{F}_{v \to e}: \mathcal{F}(v) \to \mathcal{F}(e)$. These maps encode the crucial \textbf{consistency conditions} of the system \cite{hansen2019gentle, hansen2021laplacians, arne2023sheaves}. They dictate how data from a vertex's private state space must be transformed, projected, or translated to be expressed and compared within the shared context of an edge.
\end{itemize}

This architecture provides a natural model for systems where an agent's internal state is distinct from what it communicates or shares with its neighbors. The vertex stalks $\mathcal{F}(v)$ represent the ``private'' data, while the edge stalks $\mathcal{F}(e)$ represent the ``public'' spaces for discourse. The restriction maps model the act of communication or expression itself \cite{hansen2021opinion}. This decoupling of internal state from communicated information is a massive leap in modeling fidelity compared to the graph Laplacian, where the node's value is precisely the value that is communicated.

\subsection{Global Sections and the 0th Cohomology Group ($H^0$)}
With the sheaf structure in place, we can define what it means for data to be consistent across the entire network.
\begin{itemize}
    \item A \textbf{section} (or 0-cochain) is an assignment of a data vector $\mathbf{x}_v \in \mathcal{F}(v)$ to each vertex in the graph. The collection of all such assignments, $\mathbf{x} = \{\mathbf{x}_v\}_{v \in V}$, forms a vector in the total space of 0-cochains, $C^0(G; \mathcal{F}) = \bigoplus_{v \in V} \mathcal{F}(v)$ \cite{hansen2019gentle, arne2023sheaves}.
    \item A \textbf{global section} is a special type of section that is globally consistent. This means that for every edge $e=(u,v)$ in the network, the data from nodes $u$ and $v$, when mapped into the edge's discourse space, must agree. This is the core consistency condition: $\mathcal{F}_{u \to e}(\mathbf{x}_u) = \mathcal{F}_{v \to e}(\mathbf{x}_v)$ \cite{ayzenberg2024sheaf, arne2023sheaves}. A global section represents a state of perfect data fusion or a sophisticated consensus across the network.
    \item The set of all global sections forms a vector space known as the \textbf{0th sheaf cohomology group}, denoted $H^0(G, \mathcal{F})$ \cite{hansen2019distributed, riess2022cellular}. This space represents the complete set of all possible stable, consistent equilibrium states for the entire network. Its dimension indicates the number of independent ``modes'' of consensus the system can support.
\end{itemize}

The abstract definition of a sheaf in mathematics is built on a fundamental ``gluing axiom,'' which states that if one has a collection of compatible data defined on small, overlapping regions, they can be uniquely ``glued'' together to form a single, consistent piece of data on the union of those regions \cite{bredon2012sheaf, riess2022cellular}. The existence of a global section in $H^0(G, \mathcal{F})$ is the computational embodiment of this gluing property for an entire network. This reveals that sheaf theory is not merely an analogy for data fusion; it is the canonical mathematical formalization of the local-to-global consistency problem that lies at the heart of data fusion and distributed consensus \cite{robinson2016sheaves, hansen2021laplacians}.

\subsection{A Concrete Example: Constructing a Sheaf for a Sensor Network}
To make these abstract concepts tangible, consider a simple sensor network where multiple sensors have overlapping fields of view \cite{hansen2019gentle}.
\begin{itemize}
    \item \textbf{Scenario:} A set of sensors are deployed to measure a physical quantity (e.g., temperature as a function of position) over a large area. Each sensor can only observe a subset of the total area, but these subsets overlap.
    \item \textbf{Graph Construction:} We build a graph $G$ where each vertex $v$ corresponds to a sensor, and an edge $e$ connects two vertices $u$ and $v$ if their fields of view, $U_u$ and $U_v$, have a non-empty intersection.
    \item \textbf{Stalk Definition:}
    \begin{itemize}
        \item The vertex stalk $\mathcal{F}(v)$ for a sensor $v$ is the vector space of all possible functions it can measure, i.e., functions defined on its field of view: $\mathcal{F}(v) = \{f: U_v \to \mathbb{R}\}$.
        \item The edge stalk $\mathcal{F}(e)$ for an edge between $u$ and $v$ is the space of functions defined on their shared domain: $\mathcal{F}(e) = \{f: U_u \cap U_v \to \mathbb{R}\}$.
    \end{itemize}
    \item \textbf{Restriction Map Definition:} The restriction maps are defined by the natural mathematical operation of function restriction. The map $\mathcal{F}_{v \to e}: \mathcal{F}(v) \to \mathcal{F}(e)$ simply takes a function $f_v$ defined on the larger domain $U_v$ and restricts its domain to the smaller, overlapping region $U_u \cap U_v$.
    \item \textbf{Meaning of a Global Section:} In this context, the consistency condition $\mathcal{F}_{u \to e}(\mathbf{x}_u) = \mathcal{F}_{v \to e}(\mathbf{x}_v)$ means that the function measured by sensor $u$ and the function measured by sensor $v$ must agree on their overlapping field of view. A global section of this sheaf---an element of $H^0(G, \mathcal{F})$---corresponds to a single, self-consistent global function defined over the entire area covered by the union of all sensor fields of view. The sheaf structure provides the precise mathematical machinery to check if the local measurements can be ``glued'' together into such a coherent global understanding.
\end{itemize}

\section{The Sheaf Laplacian: Encoding Complex Relational Structure}
The sheaf provides the static data structure for modeling complex relationships. To study the dynamics and global properties of signals on this structure, we need an operator analogous to the graph Laplacian. This operator is the \textbf{sheaf Laplacian}, a powerful matrix that fully captures the rich relational information encoded in the sheaf.

\subsection{Formal Construction from the Coboundary Operator}
The sheaf Laplacian is constructed from a more fundamental linear map called the coboundary operator.
\begin{enumerate}
    \item \textbf{Cochain Spaces:} As previously defined, we have the space of \textbf{0-cochains}, $C^0(G; \mathcal{F}) = \bigoplus_{v \in V} \mathcal{F}(v)$, which represents all possible assignments of data to the vertices (i.e., signals on the network). We also have the space of \textbf{1-cochains}, $C^1(G; \mathcal{F}) = \bigoplus_{e \in E} \mathcal{F}(e)$, which represents all possible assignments of data to the edges \cite{hansen2019gentle}.
    \item \textbf{Coboundary Operator ($\delta$):} The \textbf{coboundary operator}, $\delta: C^0(G; \mathcal{F}) \to C^1(G; \mathcal{F})$, is a linear map that measures the local disagreement or inconsistency of a signal. To define it, we first assign an arbitrary orientation to each edge (e.g., for an edge between $u$ and $v$, we designate it as $e=u \to v$). For a given signal $\mathbf{x} = \{\mathbf{x}_v\} \in C^0(G; \mathcal{F})$, the value of its coboundary on the edge $e$, denoted $(\delta \mathbf{x})_e$, is the difference between the data from the two endpoints, after being mapped into the edge's discourse space \cite{hansen2019gentle, arne2023sheaves}:
    $$(\delta \mathbf{x})_e = \mathcal{F}_{v \to e}(\mathbf{x}_v) - \mathcal{F}_{u \to e}(\mathbf{x}_u)$$ 
    The resulting vector $\delta \mathbf{x} \in C^1(G; \mathcal{F})$ is a collection of these local disagreement values, one for each edge. A signal $\mathbf{x}$ is a global section if and only if its disagreement is zero on every edge, which means $\delta \mathbf{x} = 0$. Therefore, the kernel of the coboundary operator is precisely the space of global sections: $\ker(\delta) = H^0(G, \mathcal{F})$ \cite{hansen2019distributed}.
    \item \textbf{Sheaf Laplacian ($L_{\mathcal{F}}$):} The sheaf Laplacian is then defined as the composition of the coboundary operator and its adjoint (transpose), $\delta^T: C^1(G; \mathcal{F}) \to C^0(G; \mathcal{F})$ \cite{ayzenberg2024sheaf, hansen2019distributed, arne2023sheaves}:
    $$L_{\mathcal{F}} = \delta^T \delta$$
    This operator $L_{\mathcal{F}}$ is a positive semi-definite linear map from the space of vertex signals back to itself, $L_{\mathcal{F}}: C^0(G; \mathcal{F}) \to C^0(G; \mathcal{F})$. Crucially, while the definition of $\delta$ depends on a choice of edge orientations, the final Laplacian $L_{\mathcal{F}}$ is independent of this choice.
\end{enumerate}

The structure of $L_{\mathcal{F}}$ is a block matrix, with blocks indexed by the vertices of the graph. If vertex $u$ has a stalk of dimension $d_u$, the block $L_{uv}$ will be a $d_u \times d_v$ matrix. The blocks are defined as follows \cite{hansen2019distributed, arne2023sheaves}:
\begin{itemize}
    \item \textbf{Diagonal blocks ($u=v$):} $L_{vv} = \sum_{e \sim v} (\mathcal{F}_{v \to e})^T (\mathcal{F}_{v \to e})$
    \item \textbf{Off-diagonal blocks ($u \neq v$, connected by edge $e$):} $L_{uv} = -(\mathcal{F}_{u \to e})^T (\mathcal{F}_{v \to e})$
\end{itemize}

The coboundary operator $\delta$ can be understood as a ``disagreement sensor.'' It takes a global assignment of states $\mathbf{x}$ and produces a vector of local tensions $\delta \mathbf{x}$. The sheaf Laplacian $L_{\mathcal{F}} = \delta^T\delta$ can then be interpreted as an operator that first measures these local tensions via $\delta$, and then aggregates the effect of these tensions back onto the nodes via $\delta^T$. A node will therefore experience a strong ``pull'' from the Laplacian if it is involved in many relationships that are in a state of high tension. This provides a clear, mechanistic intuition for the consensus dynamics it generates.

\subsection{A True Generalization}
The sheaf Laplacian is not merely an alternative to the graph Laplacian; it is a vast generalization that contains the graph Laplacian and its variants as special cases \cite{ayzenberg2024sheaf, hansen2019distributed, arne2023sheaves}.
\begin{itemize}
    \item If we define a \textbf{constant sheaf} where all stalks $\mathcal{F}(v)$ and $\mathcal{F}(e)$ are the one-dimensional space $\mathbb{R}$ and all restriction maps are the identity function, the sheaf Laplacian $L_{\mathcal{F}}$ reduces exactly to the standard graph Laplacian $L$.
    \item If the stalks are $\mathbb{R}$ but the restriction maps are defined as $\sqrt{w_{uv}}$, $L_{\mathcal{F}}$ becomes the \textbf{weighted graph Laplacian}.
    \item If the stalks are a higher-dimensional space like $\mathbb{R}^n$ and the restriction maps include rotations or sign changes (e.g., multiplication by -1), $L_{\mathcal{F}}$ can represent \textbf{connection Laplacians} and the Laplacians of \textbf{signed graphs}.
\end{itemize}

\subsection{The Power of Expressivity}
This generalized structure allows the sheaf Laplacian to model a rich variety of phenomena that are inaccessible to standard graph-based methods \cite{cassara2024cellular, hansen2021laplacians}.
\begin{itemize}
    \item \textbf{Heterogeneous Data:} Nodes can have stalks of different dimensions, natively representing networks of agents with different roles and capabilities (e.g., a sensor with a 3-dimensional vector output communicating with a sensor that only measures scalar temperature) \cite{cassara2024cellular, schaub2024sheaf}.
    \item \textbf{Asymmetric and Directed Interactions:} The restriction maps $\mathcal{F}_{u \to e}$ and $\mathcal{F}_{v \to e}$ are independent, allowing the model to capture non-reciprocal or directed influence between nodes.
    \item \textbf{Arbitrary Linear Constraints:} The framework is general enough to model any system of local linear constraints, making it a powerful tool for analyzing physical, engineering, and control systems \cite{hansen2021laplacians}.
\end{itemize}

Unlike the standard graph Laplacian, which is a purely structural operator, the sheaf Laplacian is fundamentally an ``information-theoretic'' operator. Its entries are compositions of the restriction maps ($F^TF$), meaning it encodes the rules of information transformation and comparison across the network. This represents a profound shift from analyzing network structure alone to analyzing the structure of information flow and compatibility \textit{on} that network.

\section{Sheaf-Theoretic Consensus and Signal Processing}
The sheaf Laplacian provides the operational core for defining and achieving sophisticated forms of consensus and for generalizing the tools of signal processing to networks with complex, structured data.

\subsection{The Kernel as the Consensus Manifold}
The most important property of the sheaf Laplacian is the identity that connects the algebraic operator to the underlying topological data structure:
$$\ker(L_{\mathcal{F}}) = H^0(G, \mathcal{F})$$ 
This result establishes that the null space of the sheaf Laplacian is precisely the space of global sections of the sheaf \cite{ayzenberg2024sheaf, hansen2019distributed, hansen2019toward}. This has a profound implication for the notion of consensus. In the classical graph Laplacian case, the kernel was the space of constant vectors, meaning consensus was uniquely defined as uniform agreement. In the sheaf-theoretic framework, the set of all possible consensus states is a vector subspace, $H^0(G, \mathcal{F})$, which can be high-dimensional and contain highly structured, non-uniform states of agreement \cite{hansen2021laplacians}.

This reframes the concept of consensus entirely. It is no longer a single destination point (e.g., the network-wide average) but a ``consensus manifold''---a subspace of valid configurations. This is far more realistic for complex systems. For example, it can model situations of stable disagreement or polarization, where different subgroups in a social network maintain distinct but internally consistent opinions, a phenomenon known as the ``community cleavage problem'' \cite{hansen2021opinion}. For a sensor network observing a complex event, there might be multiple global interpretations of the local data that are all self-consistent; the space $H^0(G, \mathcal{F})$ captures this entire space of possibilities.

\subsection{Sheaf Diffusion Dynamics}
This understanding of consensus leads directly to a practical, distributed algorithm for data fusion. This algorithm is based on the \textbf{sheaf heat equation}, a generalization of the graph diffusion process \cite{hansen2021opinion, hansen2019distributed}:
$$\frac{d\mathbf{x}}{dt} = -\alpha L_{\mathcal{F}} \mathbf{x}$$
where $\mathbf{x}(t) \in C^0(G, \mathcal{F})$ is the vector of data at all nodes at time $t$, and $\alpha > 0$ is a rate constant. This equation describes a diffusion process where each node's state evolves based on its ``tension'' with its neighbors, as measured by the sheaf Laplacian. The dynamics of this system have a guaranteed convergence property: any initial state $\mathbf{x}(0)$ (e.g., a set of raw, noisy sensor readings) will evolve over time and asymptotically converge to the orthogonal projection of $\mathbf{x}(0)$ onto the consensus subspace $H^0(G, \mathcal{F})$ \cite{hansen2019distributed}. In other words, the network will naturally ``relax'' from a state of inconsistency into the closest possible globally consistent state. This provides a powerful and principled method for data fusion: initialize the network with the observed local data and let it evolve according to these local dynamics. The final equilibrium state is the optimal, fused global estimate that best agrees with the initial measurements while satisfying all of the network's relational constraints \cite{hansen2019learning, purvine2018sheaf}.

This diffusion process can be understood as a form of ``semantically-aware'' local averaging. Unlike standard graph diffusion which simply mixes scalar values, sheaf diffusion mixes data only after it has been transformed by the restriction maps into a common ``language'' or frame of reference within the edge stalks. The correction applied to a node's state is based on a disagreement computed in this shared context. This is what allows the process to handle complex, heterogeneous relationships and is the key to its effectiveness in applications like Sheaf Neural Networks for heterophilic graphs \cite{barbero2022attention}.

\subsection{Signal Filtering and Processing on Sheaves}
The sheaf Laplacian framework extends the entire field of Graph Signal Processing (GSP) to handle more complex data and network structures. In GSP, the graph Laplacian is treated as a ``shift operator,'' and its eigenbasis provides a ``graph Fourier transform'' for analyzing the frequency content of signals on a graph. The sheaf Laplacian and its associated diffusion operators (e.g., $D_{\mathcal{F}} = I - \alpha L_{\mathcal{F}}$) serve as generalized shift operators for signals in the structured space $C^0(G, \mathcal{F})$ \cite{ayzenberg2024sheaf, arne2023sheaves}.
\begin{itemize}
    \item \textbf{Spectral Analysis:} The eigenbasis of $L_{\mathcal{F}}$ provides a ``sheaf Fourier basis,'' allowing for the spectral decomposition of signals defined by the sheaf. The eigenvalues correspond to generalized frequencies, and the eigenvectors (or ``sheaf harmonics'') are the fundamental modes of variation that respect the sheaf's structure \cite{arne2023sheaves, hansen2019toward}.
    \item \textbf{Sheaf Convolution:} Just as in GSP, convolution with a filter can be defined as multiplication in the spectral domain. To avoid the high computational cost of eigendecomposition, these convolutional filters can be parameterized as learnable polynomials in the sheaf diffusion operator, e.g., $\mathbf{y}_{out} = P(D_{\mathcal{F}})\mathbf{x}_{in}$. This approach, particularly using low-degree polynomials, forms the architectural foundation of Sheaf Neural Networks, enabling deep learning on data with complex relational structure \cite{ayzenberg2024sheaf}.
\end{itemize}

\section{Advanced Application: Quantifying and Detecting Network Anomalies}
One of the most powerful applications of the sheaf Laplacian framework is in the detection of anomalies, faults, and inconsistencies within a network. It provides a principled, model-based approach that redefines an anomaly not as a statistical outlier, but as a violation of the network's fundamental relational structure.

\subsection{Dirichlet Energy as a Measure of Inconsistency}
The quadratic form associated with the sheaf Laplacian is known as the \textbf{Dirichlet Energy} of a signal $\mathbf{x} \in C^0(G, \mathcal{F})$ \cite{ayzenberg2024sheaf}. It can be expressed as the squared norm of the coboundary of the signal:
$$E(\mathbf{x}) = \mathbf{x}^T L_{\mathcal{F}} \mathbf{x} = \mathbf{x}^T \delta^T \delta \mathbf{x} = ||\delta \mathbf{x}||^2 = \sum_{e \in E} ||(\delta \mathbf{x})_e||^2 = \sum_{e=(u,v) \in E} ||\mathcal{F}_{v \to e}(\mathbf{x}_v) - \mathcal{F}_{u \to e}(\mathbf{x}_u)||^2$$
This equation has a clear and powerful interpretation: the total Dirichlet energy of the network is the sum of the squared magnitudes of the local disagreements on every single edge \cite{mahdi2024moving}. A signal is considered ``smooth'' or consistent with the sheaf structure if its Dirichlet energy is low \cite{arne2023sheaves}. Conversely, a signal that is highly inconsistent or anomalous will generate significant local tensions, resulting in a high Dirichlet energy.

This moves anomaly detection away from simply identifying unusual values and towards identifying ``broken relationships.'' A sensor's reading might be within its normal range, but if it is inconsistent with the readings of its neighbors according to the rules defined by the sheaf's restriction maps, it will contribute to a high Dirichlet energy and be flagged as anomalous. This is a far more sophisticated, context-aware form of anomaly detection, which is crucial for monitoring the health of complex, interacting systems \cite{chen2024graph}.

\subsection{Spectral Anomaly Detection}
The spectrum of the sheaf Laplacian provides a ``dictionary of disagreements'' that can be used for highly sensitive anomaly detection. The eigenvectors of $L_{\mathcal{F}}$ represent the fundamental modes of variation, or ``sheaf harmonics,'' of the network.
\begin{itemize}
    \item Eigenvectors associated with small, non-zero eigenvalues correspond to ``low-energy'' or ``low-frequency'' modes of variation. These are patterns of deviation from perfect consensus that the network can support without creating much internal tension.
    \item Eigenvectors associated with large eigenvalues correspond to ``high-energy'' or ``high-frequency'' modes. These are patterns of node values that are maximally at odds with the sheaf's consistency rules \cite{hansen2019toward}.
\end{itemize}

This provides a principled framework for spectral anomaly detection. An observed signal $\mathbf{x}$ from the network can be projected onto the eigenbasis of $L_{\mathcal{F}}$. If the signal has large coefficients corresponding to the high-eigenvalue modes, it means the signal contains patterns of variation that are highly inconsistent with the network's expected behavior, thus indicating an anomaly. This allows for not only detecting that an anomaly exists but also diagnosing its nature by identifying which modes of disagreement are most active.

\subsection{Use Case: Detecting a Faulty Sensor}
Consider a sensor network operating normally, with low overall Dirichlet energy. If one sensor becomes faulty and begins to transmit erroneous data, $\mathbf{x}_{faulty}$, this value will likely be highly inconsistent with the readings of its neighbors, $\mathbf{x}_{neighbor}$. This will create large local disagreement terms, $||\mathcal{F}_{faulty \to e}(\mathbf{x}_{faulty}) - \mathcal{F}_{neighbor \to e}(\mathbf{x}_{neighbor})||^2$, on all edges connected to the faulty sensor. These local tensions will cause a sharp increase in the network's total Dirichlet energy, signaling an anomalous event. Furthermore, by analyzing which nodes and edges are the primary contributors to this energy spike, the specific faulty node can be precisely located \cite{chen2024graph}. This method flags violations of the network's ``relational contract,'' providing a robust alternative to purely statistical outlier detection.

\section{A Comparative Analysis and Practical Considerations}
While the sheaf Laplacian framework offers a significant leap in modeling power, its adoption requires a careful consideration of its advantages and limitations compared to established methods.

\subsection{Sheaf-Theoretic vs. Traditional Frameworks}
The following table provides a structured comparison of the sheaf Laplacian framework against the traditional approaches of gossip algorithms (underpinned by the graph Laplacian) and Kalman filters, synthesizing the key trade-offs discussed throughout this report \cite{purvine2018sheaf}.

\begin{table}[h!]
\centering
\caption{Comparison of Distributed Data Integration Frameworks}
\label{tab:comparison}
\small
\begin{tabular}{p{0.14\linewidth} p{0.24\linewidth} p{0.24\linewidth} p{0.24\linewidth}}
\toprule
\textbf{Feature} & \textbf{Standard Gossip} & \textbf{Kalman Filter} & \textbf{Sheaf Laplacian} \\
\midrule
\textbf{Data Type} & Primarily scalar values & Vector-valued states & Arbitrary vector spaces, functions \cite{hansen2021laplacians, arne2023sheaves} \\
\textbf{Complexity} & Pairwise adjacency connections & Linear state-space; encoded in system matrices & Arbitrary transformations via restriction maps \cite{hansen2021laplacians, hansen2021opinion} \\
\textbf{Consensus} & Convergence to a single scalar average \cite{xiao2005local} & Convergence to true state trajectory & Convergence to harmonic consensus manifold \cite{hansen2019distributed, hansen2021opinion} \\
\textbf{Topology} & Decentralized, local exchanges & Often centralized or complex distributed & Inherently decentralized and local \cite{hansen2021laplacians} \\
\textbf{Heterogeneity} & Limited & Common state space assumption & Native support for varied role spaces \cite{cassara2024cellular, schaub2024sheaf} \\
\textbf{Abstraction} & Low & Medium & High \cite{hansen2021laplacians, ayzenberg2024sheaf, bredon2012sheaf} \\
\textbf{Limitations} & Low expressivity \cite{hansen2021laplacians} & Model mismatch vulnerability & High conceptual/computational costs \cite{numberanalytics2025limitations, sardellitti2024learning} \\
\bottomrule
\end{tabular}
\end{table}

\subsection{Advantages of the Sheaf Framework}
The primary advantages of the sheaf-based approach are its unparalleled expressivity and its unifying theoretical foundation.
\begin{itemize}
    \item \textbf{Unmatched Expressivity:} The framework can model complex, heterogeneous, and context-dependent interactions that are far beyond the reach of any other single framework. It allows for the representation of systems with agents of varying capabilities, asymmetric influence, and sophisticated, non-uniform agreement states \cite{cassara2024cellular, barbero2022sheaf}.
    \item \textbf{Unified Theory:} Sheaf theory provides a single, canonical mathematical language for a diverse set of problems in signal processing, control theory, and distributed computing that were previously addressed by a disparate collection of ad-hoc tools \cite{hansen2021laplacians, purvine2018sheaf}. It reveals the deep connections between these domains.
    \item \textbf{Principled Handling of Inconsistency:} The framework moves beyond simple error handling to a formal, quantitative measure of data inconsistency through the Dirichlet energy and a characterization of systemic obstructions to agreement through the lens of sheaf cohomology \cite{hansen2019learning, ayzenberg2024sheaf}.
\end{itemize}

\subsection{Challenges and Limitations}
Despite its power, the sheaf framework presents significant practical challenges that must be acknowledged.
\begin{itemize}
    \item \textbf{Conceptual and Mathematical Complexity:} The theory is rooted in abstract mathematics, including category theory and algebraic topology. This represents a steep learning curve and a significant barrier to entry for many engineers and practitioners not trained in these fields \cite{hansen2021laplacians, ayzenberg2024sheaf}.
    \item \textbf{The ``Sheaf Learning'' Problem:} The effectiveness of the sheaf Laplacian is entirely dependent on the definition of the underlying sheaf---specifically, the restriction maps. Manually designing these maps requires deep and precise domain knowledge. The problem of \textit{learning} the optimal sheaf structure directly from data is a critical, and still very active, area of research. Without robust methods for sheaf learning, the practical application of the framework can be limited \cite{hansen2019learning, sardellitti2024learning}.
    \item \textbf{Computational Cost:} The sheaf Laplacian is a block-structured matrix, and its size grows with the sum of the dimensions of the vertex stalks. Constructing this matrix and performing operations on it, such as eigendecomposition, can be significantly more computationally intensive than for the sparse, scalar-entry graph Laplacian, especially for networks with many nodes or high-dimensional stalks \cite{numberanalytics2025limitations}.
\end{itemize}

\section{Future Horizons in Network Science and Signal Processing}
The theory of the sheaf Laplacian is not a static endpoint but an active and rapidly evolving field of research. Its integration with machine learning and its application to next-generation network challenges are charting exciting future directions.

\subsection{Sheaf Learning and Automated Model Discovery}
Perhaps the most critical frontier for the practical application of sheaf theory is \textbf{sheaf learning}. This line of research aims to develop algorithms that can automatically infer the optimal sheaf structure---specifically, the restriction maps---directly from observed data \cite{hansen2019learning, sardellitti2024learning}. The typical approach involves formulating an optimization problem to find the set of restriction maps that minimizes the total Dirichlet energy (or ``smoothness'') of a given set of training signals. This has the potential to transform sheaf theory from a purely descriptive modeling tool into a powerful discovery tool. By inspecting the learned restriction maps, researchers may be able to uncover previously unknown relational structures, causal links, or effective interaction mechanisms hidden within complex datasets \cite{mahdi2024moving}.

\subsection{Sheaf Neural Networks (SNNs)}
The sheaf Laplacian and its associated diffusion operators are being integrated as layers within Graph Neural Network (GNN) architectures, giving rise to \textbf{Sheaf Neural Networks (SNNs)} \cite{bodnar2022neural, barbero2022sheaf}. This development is significant because SNNs have been shown to overcome major pathological issues of traditional GNNs. For instance, standard GNNs often perform poorly on \textbf{heterophilic graphs}, where connected nodes tend to be of different classes. The simple message-passing of GNNs struggles in these cases, but the structured, semantically-aware diffusion of SNNs can effectively model these complex relationships. SNNs have also shown promise in mitigating the problem of \textbf{oversmoothing}, where node representations become indistinguishable after many layers of message passing \cite{barbero2022attention}. Research is already pushing beyond basic SNNs to more advanced architectures like Sheaf Attention Networks and their application to higher-order structures like hypergraphs \cite{bodnar2022neural, barbero2022sheaf, barbero2022attention}.

\subsection{Emerging Applications}
The expressive power of the sheaf framework is enabling novel solutions in a growing number of domains:
\begin{itemize}
    \item \textbf{Federated and Multi-Task Learning:} Sheaves provide a natural framework for decentralized machine learning, where different clients may have heterogeneous data or be training different but related models. A sheaf can model the relationships between client tasks, and the sheaf Laplacian can be used as a regularization term in the optimization objective to encourage related clients to learn similar model parameters, improving performance and communication efficiency \cite{schaub2024sheaf}.
    \item \textbf{Multi-Agent Robotics and Control:} In robotics, consensus often means achieving a complex physical formation or a coordinated action rather than simply agreeing on a scalar value. Sheaves are being used to design sophisticated coordination and control algorithms for heterogeneous teams of robots and drones, where the sheaf structure encodes the desired geometric or functional relationships between agents \cite{cassara2024cellular, barbero2022sheaf}.
    \item \textbf{Delay-Tolerant Networking (DTN):} For future communication networks characterized by extremely long delays and intermittent connectivity, such as a potential Solar System Internet, traditional routing protocols fail. Because of its powerful and robust local-to-global framework, sheaf theory is being explored as a more suitable foundational theory for routing and information flow in these challenging environments \cite{ciucci2021sheaves}.
    \item \textbf{Generalization to Abstract Structures:} Research is actively extending these concepts from simple graphs to more general combinatorial structures like cell complexes, hypergraphs, and partially ordered sets (posets) \cite{ghrist2011computational, ayzenberg2024sheaf}. This will further broaden the applicability of sheaf-theoretic tools to a wider range of problems in data science and beyond.
\end{itemize}

The convergence of these research threads points toward a future of truly adaptive, self-organizing distributed systems. One can envision a feedback loop where a system observes its own behavior, \textit{learns} the rules of its own internal consistency by optimizing a sheaf, and then \textit{acts} on those learned rules using a sheaf-based controller or neural network. This represents a blueprint for systems that can adapt their fundamental operating logic in response to a changing environment---a significant step beyond systems that merely adapt parameters within a fixed logic. The sheaf framework can be viewed as a principled method to ``compile'' a set of high-level desired consistency relationships into a set of local, distributed dynamic rules that will provably enforce them, bridging the gap between system specification and implementation.

\section{Conclusion}
This report has traced the intellectual journey from the simple, scalar world of the graph Laplacian to the rich, structured, and heterogeneous domain of the sheaf Laplacian. The analysis demonstrates that this evolution is not merely a technical upgrade but a fundamental paradigm shift in how we model, analyze, and engineer distributed systems. The graph Laplacian, while elegant, constrains our view of network dynamics to simple connectivity and uniform consensus. The sheaf Laplacian shatters these constraints, providing a language capable of describing the complex, context-dependent, and multi-dimensional relationships that define modern sensor networks, robotic teams, and decentralized AI.

By associating structured data with the very fabric of the network---the nodes and edges---and by encoding the rules of interaction as linear maps, the sheaf-theoretic framework offers a canonical and unified approach to the core challenges of data fusion and consensus. It redefines consensus not as a single point of agreement but as a manifold of valid, globally consistent states. It provides a principled, distributed dynamic---sheaf diffusion---that allows a network to naturally relax into the optimal state of agreement. And it transforms anomaly detection from a search for statistical outliers into a context-aware process of identifying broken relational contracts.

While the mathematical sophistication of sheaf theory presents a barrier to entry, and the challenge of learning sheaf structures from data remains an active frontier of research, the potential benefits are immense. The sheaf Laplacian stands as a foundational operator in modern network science, providing the essential mathematical rigor to tame the complexity of emerging distributed systems. It builds a powerful bridge between topology, geometry, and signal processing, offering a profound lens through which to understand and engineer the collective intelligence of the future.

\bibliographystyle{plain}
\bibliography{references}

\end{document}